\documentclass[sigconf,screen,pbalance=true]{acmart}

\AtBeginDocument{%
	}

\usepackage{booktabs}
\usepackage{amsmath}
\usepackage{xurl}
\usepackage{graphicx}
\usepackage{tikz}
\usetikzlibrary{arrows.meta,backgrounds,fit,positioning,shapes.geometric}
\usepackage{microtype}
\setcopyright{cc}
\setcctype{by}
\acmDOI{10.1145/3843282.3844422}
\acmYear{2026}
\copyrightyear{2026}
\acmISBN{979-8-4007-2985-0/2026/10}
\acmConference[AgenticDev '26]{Proceedings of the 1st International Workshop on Agentic AI for Next-Generation Software Development}{October 12--16, 2026}{Munich, Germany}
\acmBooktitle{Proceedings of the 1st International Workshop on Agentic AI for Next-Generation Software Development (AgenticDev '26), October 12--16, 2026, Munich, Germany}
\acmSubmissionID{asews26agenticdevmain-p18-p}
\received{2026-07-27}
\received[accepted]{2026-08-20}

\begin{document}
		
	\title{Cost-Effective Repository Exploration for Agentic Issue Localization}

	\author{Mohammad Nour Al Awad}
	\correspondingauthor
	\orcid{0009-0008-5078-9800}
	\affiliation{%
		\institution{ITMO University}
		\city{Saint Petersburg}
		\country{Russian Federation}
	}
	\email{mohammadnouralawad@itmo.ru}
	
	\author{Sergey Ivanov}
	\orcid{0000-0002-1128-2942}
	\affiliation{%
		\institution{ITMO University}
		\city{Saint Petersburg}
		\country{Russian Federation}
	}
	\email{svivanov@itmo.ru}
	
	\begin{abstract}
		Repository exploration is a distinct and costly stage of coding-agent pipelines: before generating a patch, an agent must identify which repository files are likely to matter. We study whether this stage can be delegated to lower-cost models while retaining useful localization quality. Using our IssueLoc-Bench, we evaluate five explorer models under the same read-only interactive interface on 499 SWE-bench Verified-derived tasks and 500 tasks from 153 additional repositories. We measure early candidate discovery, top-three gold-file coverage, strict file-set recovery, agent time, and token usage, with paired instance-level uncertainty and repository-clustered sensitivity analysis. The highest-quality explorer leads across localization metrics, but substantially cheaper operating points emerge: depending on the model and evaluation arm, lower-cost explorers retain approximately 78--94\% of the reference Hit@3 and 73--92\% of its F1 while reducing mean agent time by 41--88\% and token usage by 84--95\%. The preferred operating point depends on how localization is consumed downstream: ranking and coverage metrics characterize recoverable candidate handoffs, whereas F1 and exact match characterize restrictive file gates. These results support treating repository exploration as an independently measurable and budgetable stage of modular coding agents, with explorer selection guided by the downstream handoff contract.
	\end{abstract}

	
	
	\maketitle
	
	\section{Introduction}
	\label{sec:intro}
	Modern coding agents are multi-stage systems: they interpret an issue, explore the repository, synthesize a patch, and validate the result. Model allocation across these stages is usually implicit because one model performs most or all of the trajectory. Repository exploration is an especially natural stage at which to make allocation explicit. It produces a compact intermediate artifact---a ranked candidate-file list---whose quality and production cost can be measured independently of patch generation.
	
	Before repair, the agent must determine \emph{where to look} by mapping a natural-language issue report to files containing relevant evidence and plausible edit locations. This repository-exploration problem is difficult because issue reports are often incomplete, repositories are large, and the connection between a problem description and concrete files is rarely explicit \cite{AutoFL2024,LLMCodeFL2025,LocAgent2025,GottaCatchEmAll2025}. Poor exploration can introduce irrelevant context, consume downstream reasoning budget, and misdirect subsequent work.

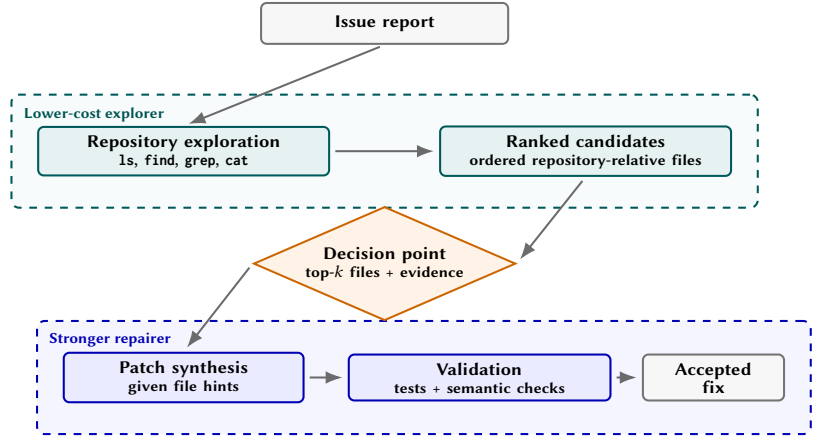
\begin{figure*}[t]
		\centering
		\resizebox{0.60\textwidth}{!}{\begin{tikzpicture}[
	x=1cm,
	y=1cm,
	font=\sffamily\footnotesize,
	>=Latex,
	node distance=0,
	basebox/.style={
		rounded corners=2.2pt,
		thick,
		align=center,
		inner xsep=4pt,
		inner ysep=3pt,
		minimum height=0.56cm
	},
	neutralbox/.style={basebox, draw=black!55, fill=black!3},
	explorerbox/.style={basebox, draw=teal!70!black, fill=teal!10},
	repairbox/.style={basebox, draw=blue!70!black, fill=blue!8},
	stagebox/.style={
		dashed,
		thick,
		rounded corners=3pt,
		inner sep=7pt
	},
	arrow/.style={->, thick, draw=black!58, shorten >=2pt, shorten <=2pt}
]

\node[neutralbox, text width=3.1cm] (issue) at (4.85,5.60)
	{\footnotesize\bfseries Issue report};

\node[explorerbox, text width=3.7cm] (explore) at (2.10,3.85)
	{\footnotesize\bfseries Repository exploration\\[-1pt]{\scriptsize \texttt{ls}, \texttt{find}, \texttt{grep}, \texttt{cat}}};
\node[explorerbox, text width=3.7cm] (rank) at (7.60,3.85)
	{\footnotesize\bfseries Ranked candidates\\[-1pt]{\scriptsize ordered repository-relative files}};

\node[
	draw=orange!80!black,
	fill=orange!10,
	diamond,
	aspect=2.3,
	thick,
	align=center,
	text width=2.3cm,
	inner xsep=1pt,
	inner ysep=1pt
] (handoff) at (4.85,2.31)
	{\footnotesize\bfseries Decision point\\[-1pt]{\scriptsize top-$k$ files + evidence}};

\node[repairbox, text width=3.0cm] (patch) at (2.10,0.75)
	{\footnotesize\bfseries Patch synthesis\\[-1pt]{\scriptsize given file hints}};
\node[repairbox, text width=3.3cm] (validate) at (6.15,0.75)
	{\footnotesize\bfseries Validation\\[-1pt]{\scriptsize tests + semantic checks}};
\node[neutralbox, text width=1.65cm] (fix) at (9.35,0.75)
	{\footnotesize\bfseries Accepted\\[-1pt]fix};

\begin{scope}[on background layer]
	\node[
		stagebox,
		draw=teal!70!black,
		fill=teal!3,
		fit=(explore)(rank),
		inner xsep=10pt,
		inner ysep=12pt,
		label={[anchor=north west, xshift=2pt, yshift=-2pt, text=teal!65!black, font=\sffamily\bfseries\scriptsize]north west:Lower-cost explorer}
	] {};
	\node[
		stagebox,
		draw=blue!70!black,
		fill=blue!4,
		fit=(patch)(validate)(fix),
		inner xsep=10pt,
		inner ysep=12pt,
		label={[anchor=north west, xshift=2pt, yshift=-2pt, text=blue!70!black, font=\sffamily\bfseries\scriptsize]north west:Stronger repairer}
	] {};
\end{scope}

\draw[arrow] (issue.south) -- (explore.north);
\draw[arrow] (explore.east) -- (rank.west);
\draw[arrow] (rank.south) -- (handoff.east);
\draw[arrow] (handoff.west) -- (patch.north);
\draw[arrow] (patch.east) -- (validate.west);
\draw[arrow] (validate.east) -- (fix.west);

\end{tikzpicture}}
		\caption{Model-allocation pipeline in which a lower-cost explorer ranks candidate files before stronger repair and validation.}
		\Description{Pipeline diagram showing an issue report flowing into a lower-cost repository explorer, then a ranked candidate file list, a decision point, and a stronger repair and validation stage that produces an accepted fix.}
		\label{fig:model-allocation-pipeline}
	\end{figure*}

	We therefore ask: \emph{under a fixed repository-interaction interface, what localization-quality and operational-cost trade-offs arise when different models serve as the dedicated explorer?} Loc2Repair provides evidence that predicted file localization can improve resolved rates across multiple repair backbones \cite{awad2026loc2repairframeworkevaluatingimpact}. We study the complementary upstream decision: which quality--cost operating points are available when producing that localization signal? This motivates the split-allocation design in Figure~\ref{fig:model-allocation-pipeline}, in which a dedicated explorer identifies candidate files before repair and validation. This study evaluates production of the localization signal; it does not directly measure the end-to-end repair effect of each explorer.

	Repository-grounded benchmarks have made issue resolution increasingly realistic. SWE-bench evaluates end-to-end resolution of real GitHub issues, SWE-agent demonstrates the value of repository interaction, and RepoBench and RepoGraph emphasize repository-level context and structure \cite{SWEbench2024,SWEagent2024,RepoBench2024,RepoGraph2025}. Yet end-to-end outcomes conflate issue understanding, exploration, localization, patch generation, execution, and retry policy \cite{Agentless2025,LocAgent2025}. They therefore do not isolate the narrower allocation question studied here: how should model choice and operational budget be assigned to exploration before repair begins?
	
	We study this question using IssueLoc-Bench as a controlled file-level localization substrate. Given an issue report and a pre-fix repository snapshot, an explorer interacts with the repository through a fixed read-only interface and returns an ordered list of repository-relative candidate files. The interface is held constant; the main experimental variable is the \emph{explorer model}. This keeps the study focused on model allocation rather than on interface or retrieval-method comparisons.
	
	File-level localization matches this upstream repository-exploration role. Before a repair model can reason about a patch, the agent must decide which files to open, search within, summarize, or pass forward as context. Because the resulting ordered paths can be scored without generating a patch, they create an observable boundary between exploration and the later stages of an agent trajectory.
	
	The paper evaluates two fixed arms. The first is a SWE-bench Verified-derived arm with 499 tasks, connecting the analysis to a widely used repair benchmark while changing the target to file-level localization. The second is an IssueLoc random-repository arm with 500 tasks, evaluating the same explorer interface on repository-disjoint issues from the IssueLoc construction pipeline. Together, the arms test whether the same model-allocation trade-off persists across substantially different repository and gold-file distributions.
	
	The controlled comparison tests whether lower-cost checkpoints occupy useful explorer operating points under the same interface. We distinguish two handoff contracts: a \emph{soft handoff}, in which ranked candidates guide subsequent inspection while repository access remains recoverable, and a \emph{hard gate}, in which the predicted file set constrains downstream access. Ranking and coverage metrics characterize the former, while F1 and exact match characterize the latter. This distinction prevents a single aggregate score from obscuring which part of localization quality a downstream design would consume.

	This boundary improves both measurement and system control. An end-to-end resolved-rate score cannot identify whether a failure originated in repository search, patch synthesis, validation, or retry policy. By contrast, a structured ranked-file artifact exposes whether the explorer found a useful entry point, covered a multi-file footprint, or calibrated the final file set. Holding the interface constant then attributes differences to complete explorer configurations rather than to different tool protocols. The same boundary also makes the stage independently budgetable: a system can choose an explorer operating point, candidate budget, and escalation condition without changing the representation passed to later stages.
	
	\subsection{Research Questions}
	\label{sec:rqs}
	We organize the empirical study around four research questions:
	\begin{enumerate}
		\item \textbf{RQ1 (Quality retention).} How much localization quality do lower-cost explorer models retain relative to the highest-performing evaluated explorer?
		\item \textbf{RQ2 (Cost reduction).} How much agent time and token use do lower-cost explorers save?
		\item \textbf{RQ3 (Candidate discovery vs. exact recovery).} How does quality retention differ between candidate discovery and strict file-set recovery across models and evaluation arms?
		\item \textbf{RQ4 (Cross-arm consistency).} Are conclusions consistent across the SWE-bench Verified-derived and random-repository arms?
	\end{enumerate}
	
	\paragraph{Contributions.}
	This paper makes four main contributions:
	\begin{itemize}
		\item We formulate \textbf{explorer-model allocation} as a design problem in modular coding agents, separating explorer choice from downstream patch synthesis and validation.
		
		\item We conduct a controlled comparison of five explorer models under a fixed read-only interactive interface on SWE-bench Verified-derived tasks and our 500-task random-repository arm.
		
		\item We characterize quality--cost operating points using early ranking, top-three gold coverage, strict set recovery, agent time, and token usage, together with paired instance-level uncertainty and repository-clustered sensitivity analysis.
		
		\item We show that explorer selection depends on the localization handoff contract: early-ranking metrics characterize recoverable candidate handoffs, while F1 and exact match characterize restrictive file gates.
	\end{itemize}
	
	The remainder of the paper reviews related work, defines the file-level exploration task, describes the two evaluation arms and fixed interactive runner, reports quality--cost results, and discusses implications and threats to validity.

	\section{Background and Related Work}
	\label{sec:related}
	
	\subsection{Issue Localization and Evaluation}
	A long line of work formulates bug or issue localization as a retrieval problem: given a natural-language report, rank source artifacts by their likelihood of requiring modification. Prior work has shown that localization quality depends on how textual relevance is combined with additional evidence such as report structure, version history, related reports, and repository organization \cite{SahaLKP2013ASE_BLUiR,WangL2014ICPC_PutTogether,WangLo2016AmalgamPlus,YeBL2014FSE_LTR,WangLL2014ICSME_CVSM}. More recent surveys show that this remains true in neural settings, where strong performance often depends on hybridization of lexical, historical, and learned signals rather than on a single modeling advance \cite{Niu2025CSUR_DLmeetsIR}. Recent work extends this perspective to LLM-based issue localization over repositories, including retrieve-and-rerank formulations, graph-guided agentic search, and multi-turn repository exploration \cite{SweRank2026,CoSIL2025,OrcaLoca2025,LocAgent2025}.
	Two lessons from this literature are directly relevant here. First, localization quality depends strongly on benchmark construction and evaluation protocol. Second, the meaning of a localization target depends on the artifact granularity and supervision semantics used to define it.
	
	Bug-localization research has repeatedly shown that benchmark design materially affects conclusions. Bench4BL highlighted reproducibility difficulties across datasets and implementations \cite{LeeKBJT2018ISSTA_Bench4BL}, and Akbar et al. showed that preprocessing and evaluation choices can substantially change both absolute performance and relative rankings among techniques \cite{AkbarK2020MSR_Ablation}. These findings motivate benchmarks with explicit supervision semantics, leakage-resistant splits, and multiple complementary metrics.
	
	For agentic repository exploration, ranking metrics and set metrics have different operational meanings. Ranking metrics indicate whether useful files are surfaced early enough to guide downstream inspection or repair. Exact set recovery and F1 measure a stricter target: how well the final predicted file set matches the historical resolution footprint. This distinction is especially important in recent localization benchmarks, where dataset scope, issue type diversity, and supervision targets vary substantially across Loc-Bench, SweLoc, MULocBench, and issue-at-large style collections \cite{LocAgent2025,SweRank2026,MULocBench2025,GottaCatchEmAll2025}.
	
	\subsection{Repository Exploration in Repair Pipelines}
	Recent benchmarks have shifted software-engineering evaluation from isolated snippets to full repositories. RepoBench studies repository-level retrieval and completion in multi-file settings \cite{RepoBench2024}. SWE-bench evaluates end-to-end resolution of real GitHub issues against executable test outcomes \cite{SWEbench2024}. SWE-agent further shows that interactive repository use can improve repository-grounded issue resolution \cite{SWEagent2024}. These benchmarks have been highly influential, but they evaluate \emph{composed} behavior: issue understanding, repository search, localization, patch generation, execution, and interaction policy all contribute to final outcomes.
	
	SWE-Explore asks how repository exploration should be benchmarked and connects fine-grained exploration quality to repair behavior \cite{SWEExplore}; SWE-Router asks when a cheap ongoing repair trajectory should be escalated \cite{SWERouter}. In contrast, we hold the repository-exploration interface fixed and directly measure how explorer model choice changes file-level candidate quality, runtime, and token use across two evaluation arms.
	
	IssueLoc-Bench complements these benchmarks by isolating file-level localization under repository-grounded pre-fix conditions while preserving the artifact structure used by modern repair systems. This supports a narrower systems question: which evaluated model should serve as a dedicated explorer?
	
	Several recent systems make localization explicit within repository-grounded repair. Agentless uses a localization--repair--validation pipeline; LocAgent, CoSIL, RGFL, and SGAgent likewise connect localization to issue resolution \cite{Agentless2025,LocAgent2025,CoSIL2025,OrcaLoca2025,RGFL2026,SGAgent2026}. Loc2Repair directly evaluates the downstream half of this pipeline: it decouples file localization from patch synthesis, applies localization across multiple repair backbones, and observes higher resolved rates \cite{awad2026loc2repairframeworkevaluatingimpact}. Given that localization is an empirically useful repair lever, IssueLoc-Bench studies its production: repository-grounded file-level candidate discovery over pre-fix snapshots with commit-linked supervision and a fixed interface. Together, the studies expose two separable pipeline decisions: how localization is consumed by repair and which model and operational budget are used to produce it.
	
	\section{Task Definition}
	\label{sec:task}
	
	\subsection{Task and Formalization}
	We formulate \emph{repository-grounded file-level issue localization} as an upstream exploration task. Given a natural-language issue report and a repository snapshot taken immediately before the historical human fix, an explorer must rank existing repository files that are plausible candidates for downstream repair. The task isolates repository exploration from patch synthesis and validation: explorers do not modify code, execute fixes, or predict files absent from the pre-fix snapshot.

	Each instance is defined by an issue-side input $I$ and a pre-fix repository snapshot $R$. In the main study, $I$ contains the issue text available in the task record: issue title and body for IssueLoc instances, and linked issue text or SWE-bench problem-statement fallback for transformed SWE-bench instances. Let $\mathcal{F}(R)$ denote the set of repository-relative file paths present in snapshot $R$. Given $(I, R)$, a system must produce an ordered list of distinct predicted files
	\[
	f(I, R) \rightarrow [p_1, p_2, \dots, p_k]
	\]
	such that each $p_j \in \mathcal{F}(R)$.
	
	The gold label set $G$ contains files modified by the historical human resolving commit after the filters in \S\ref{sec:data}; it operationalizes a reproducible historical resolution footprint rather than a minimal causal file set. The task is file-level, repository-grounded, pre-fix, and localization-only: outputs are existing repository-relative paths, and systems neither modify code nor validate a repair.
	
	Function- or line-level localization can be useful after a repair model has selected a candidate region, but file-level prediction is the natural operational unit for repository-scale exploration and context selection. It determines which files an agent reads, searches, summarizes, ranks, or passes to downstream synthesis and validation stages.
	
	\subsection{Interactive Explorer and Handoff}
	\label{sec:task-agent}
	The evaluated setting is interactive repository exploration. An agent may inspect the pre-fix repository snapshot through read-only exploration before emitting a final ranked file list.
	
	Interaction is restricted to repository inspection. Agents may browse directories, search for identifiers or strings, and inspect candidate files, but they may not modify repository contents. Each run must terminate with a final structured prediction containing an ordered list of repository-relative file paths. The runner records interaction traces, command counts, elapsed time, and token usage when available.
	
	This design fixes the agent interface while varying the explorer model. We evaluate both a \emph{soft handoff}, where ranking and coverage measure whether relevant files surface early for downstream inspection, and a \emph{hard gate}, where F1 and exact match measure how well the emitted set constrains downstream access.

	The distinction is operational, not merely metric terminology. Under a recoverable soft handoff, a false positive consumes inspection budget, but a repair stage that retains repository access can still follow imports, symbols, tests, or newly discovered evidence. Missing every relevant file near the top then provides a weak starting signal. Under a hard gate, an omitted file becomes inaccessible downstream, while an unnecessary file enlarges the permitted context; set completeness and calibration consequently matter more. The same ranked output can therefore have different value under the two contracts, so explorer selection should be tied to how the downstream stage uses the prediction.

	\section{Benchmark Construction and Evaluation Arms}
	\label{sec:data}
	
	IssueLoc-Bench is a benchmark artifact contributed by this work, combining task and label manifests, deterministic pre-fix snapshots, a shared evaluator, and runnable explorer pipelines. Its two arms connect a standard repair benchmark to our independently collected repository sample.
	
	\subsection{IssueLoc Collection and Manifest Design}
	\label{sec:data-manifest}
	
	We construct the random-repository arm from recent issue-resolution activity in popular, actively maintained public GitHub repositories. Within this source population, the collection samples issue--pull-request pairs and retains only merged pull requests that GitHub links to exactly one issue and that contain exactly one commit. The one-to-one linkage removes ambiguous multi-issue resolutions, while the single-commit constraint provides an unambiguous resolving revision from which to derive file labels. Repositories are separated across the released IssueLoc splits.

	Recency mitigates contamination risk because newer issue--pull-request pairs have had less opportunity to enter pretraining corpora than long-established benchmark instances. It cannot guarantee non-exposure, especially for public, popular repositories; possible training exposure therefore remains a validity threat.

	Each retained instance has task and label records joined by \texttt{instance\_id}. Task records contain the issue title, body, available discussion, repository and pull-request identifiers, split, and snapshot locators. Label records contain the resolving commit and the filtered gold file set as repository-relative paths. These labels represent the historical resolution footprint, not a minimal causal file set.
	
	\label{sec:data-materialization}
	
	The released builder fetches each resolving commit and materializes its parent tree as the explorer workspace, reducing direct post-fix leakage. Together, the public manifest, builder, evaluator, and scoring code make the inputs, labels, and reconstruction procedure inspectable and reproducible.
	
	\label{sec:data-filtering}
	
	The code-focused view excludes instances whose historical change would require predicting newly added or removed files, binary-like files, empty gold sets after filtering, or entirely markdown/text-like targets. These filters keep the target aligned with localization of existing implementation artifacts.
	
	\subsection{Evaluation Arms and Characteristics}
	\label{sec:data-splits}
	
	The empirical study uses two fixed evaluation arms, summarized in Table~\ref{tab:evaluation-arms}. The SWE-bench Verified-derived arm comes from the \texttt{test} split of \texttt{SWE-bench/SWE-bench\_Verified}; it materializes snapshots from SWE-bench base commits and derives file-level labels from patch decomposition. Of 500 source instances, 499 remain after filtering. Our IssueLoc arm comprises 500 random-repository test instances from 153 repositories, with repositories kept disjoint across IssueLoc splits. We use both arms as evaluation sets, not as train/development/test stages for model selection.
	
	\begin{table*}[t]
		\caption{Summary of the two evaluation arms. Repository size is the number of paths in the materialized pre-fix snapshot.}
		\label{tab:evaluation-arms}
		\centering
		\small
		\setlength{\tabcolsep}{5pt}
		\begin{tabular}{lrrrr}
			\toprule
			Arm & Inst. & Repos & Gold avg & Repo avg \\
			\midrule
			SWE-bench Verified-derived & 499 & 12 & 1.24 & 3898.5 \\
			IssueLoc random-repository & 500 & 153 & 2.27 & 614.2 \\
			\bottomrule
		\end{tabular}
	\end{table*}
	
	\label{sec:data-stats}
	
	The two arms differ in useful ways for model-allocation analysis. The SWE-bench Verified-derived arm has fewer gold files per task on average but much larger repositories, making candidate discovery a broad search problem. The random-repository arm has smaller repositories on average but a larger historical file footprint per task. These differences make the arms complementary: one connects to a widely used repair benchmark, and the other evaluates repository-disjoint issue localization under the IssueLoc construction pipeline.

	We report the arms separately rather than pooling their instances. A pooled average would mix two different deployment questions and would be dominated by whichever composition happened to receive more weight. Separate results reveal whether a model is consistently useful when exploration requires searching a large tree, when the historical resolution spans several files, or both. They also make the unit of generalization explicit: the SWE-derived arm supplies many tasks from a small group of repositories, whereas the random-repository arm distributes tasks across a much broader repository sample. This difference motivates both instance-level and repository-clustered uncertainty analysis.
	
	\section{Evaluation Protocol}
	\label{sec:eval}
	
	\subsection{Discovery and Set-Recovery Metrics}
	For instance $i$, let $G_i$ denote the gold file set and let $\pi_i = [p_{i1}, p_{i2}, \dots]$ be the ranked prediction list. We make ranking metrics primary because an explorer is most useful when it surfaces relevant candidate files early for downstream repair. We define
	\[
	\mathrm{Hit@}k_i = \mathbf{1}\left[\exists j \le k: p_{ij}\in G_i\right],
	\]
	and
	\[
	\mathrm{RR}_i =
	\begin{cases}
		\frac{1}{\min\{j \mid p_{ij}\in G_i\}} & \text{if such } j \text{ exists}\\
		0 & \text{otherwise.}
	\end{cases}
	\]
	We report $\mathrm{Hit@1}$, $\mathrm{Hit@3}$, $\mathrm{Hit@5}$, and mean reciprocal rank (MRR). We also report gold-file recall@3 (R@3). Let $P_i^{(3)}$ be the distinct paths among the first three predictions:
	\[
	\mathrm{R@3}_i=\frac{|P_i^{(3)}\cap G_i|}{|G_i|}.
	\]
	The reported R@3 is the macro average over instances. Hit@3 records whether \emph{any} gold file is found, whereas R@3 measures the fraction of the gold set covered within the same candidate budget. Together with MRR, these metrics capture candidate-discovery utility even when the final list is incomplete or overinclusive. Because prediction lists are variable-length, metrics at larger $k$ need not differ when an explorer emits fewer than $k$ candidates.
	
	Let $P_i$ be the deduplicated predicted file set derived from the ranked output. Define
	\[
	\mathrm{TP}_i = |P_i \cap G_i|,\quad
	\mathrm{FP}_i = |P_i \setminus G_i|,\quad
	\mathrm{FN}_i = |G_i \setminus P_i|.
	\]
	Per-instance set metrics are
	\[
	\mathrm{Prec}_i = \frac{\mathrm{TP}_i}{\mathrm{TP}_i+\mathrm{FP}_i},\quad
	\mathrm{Rec}_i = \frac{\mathrm{TP}_i}{\mathrm{TP}_i+\mathrm{FN}_i},
	\]
	\[
	\mathrm{F1}_i = \frac{2\cdot \mathrm{Prec}_i \cdot \mathrm{Rec}_i}{\mathrm{Prec}_i+\mathrm{Rec}_i},
	\quad
	\mathrm{EM}_i = \mathbf{1}[P_i = G_i].
	\]
	For an empty predicted set, we define per-instance precision as zero; whenever
	$\mathrm{Prec}_i+\mathrm{Rec}_i=0$, we define $\mathrm{F1}_i=0$, consistent with the evaluator.
	
	We report macro-averaged set metrics across scored instances $\mathcal{I}$:
	\[
	\mathrm{F1}_{\mathrm{macro}}=\frac{1}{|\mathcal{I}|}\sum_{i\in\mathcal{I}}\mathrm{F1}_i,\quad
	\mathrm{EM}=\frac{1}{|\mathcal{I}|}\sum_{i\in\mathcal{I}}\mathrm{EM}_i.
	\]
	
	These set metrics capture \emph{final prediction calibration}. Exact match is intentionally strict: it measures whether the emitted file set exactly matches the historical resolution footprint under the benchmark definition. Macro F1 is less brittle and reflects partial recovery of the gold set while penalizing overprediction.
	
	\subsection{Scoring and Operational Metrics}
	The evaluator also reports execution and validity statistics, including attempted rows, missing-label rows, error rows, error rate, and non-empty prediction rate. In our benchmark setting, an \emph{error row} is an instance for which the runner does not produce a valid final prediction payload within the allowed execution constraints, most commonly because the trajectory times out. Thus, ErrorRate should be interpreted as the fraction of attempted instances that fail to complete with a valid prediction under the benchmark runtime rules, rather than as a model-internal decoding error alone.
	\[
	\begin{aligned}
		\mathrm{ErrorRate}&=\frac{N_{\mathrm{error}}}{N_{\mathrm{attempted}}},\\
		\mathrm{NonEmptyRate}&=\frac{N_{\mathrm{pred\_nonempty}}}{N_{\mathrm{scored}}}.
	\end{aligned}
	\]
	These statistics expose whether an operating point completes reliably.
	
	Scoring applies to every recorded prediction row with a matching label; rows without matching labels are excluded. All such labeled rows remain in the denominators for Hit@k, MRR, macro F1, and exact match, including failed/error runs. If a failed run has no valid prediction, its prediction list is empty and receives zero localization credit. A valid but empty prediction likewise receives zero localization credit but is not counted as an execution error. If a predicted path does not exist in the materialized repository snapshot, it is retained and counted as a false positive under set-based metrics.
	
	The evaluator does not impose a post-hoc list-length cap; it scores the emitted prediction list as produced by the runner or model. Schema violations and timeouts are counted in error statistics but do not remove labeled rows from quality scoring.
	
	From runner traces and per-instance metadata, we aggregate average agent steps, reads, writes, shell commands, agent time, token usage, and output size. Agent time is the mean per-attempt wall-clock agent elapsed time; it includes model/API waiting, repository-tool execution, and timeout penalties. Average token counts are computed only over instances for which token-usage metadata was recorded; usage coverage can differ across model runs because failed trajectories do not always report token counts. Reported reductions are observed explorer operating-point differences, not intrinsic model-speed differences: the complete operating point includes model behavior, interaction turns, repository-tool use, stopping behavior, and serving/runtime characteristics. In IssueLoc-Bench, localization is both an accuracy task and an operational control point. Accordingly, a setting that improves exact recovery at materially higher agent time or token use should be interpreted differently from one that offers a lower-cost trajectory with weaker final set recovery.
	
	\subsection{Paired Uncertainty Quantification}
	For each arm, we align every alternative explorer with the highest-quality reference explorer by instance and define each difference as alternative minus reference. We reconstruct Hit@1, Hit@3, Hit@5, R@3, MRR, per-instance F1, exact match, agent time, and input, output, and total tokens directly from the run artifacts using the evaluator's scoring and markdown-filtering functions. We use a percentile paired bootstrap with 10,000 resamples, fixed seed 1729, and 95\% marginal confidence intervals. Quality metrics use all 499 or 500 aligned instances; operational metrics use pairwise-complete observations, with the exact sample size reported in the artifact. For binary Hit@$k$ and exact-match outcomes, we additionally report two-sided exact McNemar tests and Holm-adjusted values across the 32 binary comparisons.
	
	Instances from the same repository need not be independent. As a sensitivity analysis, we therefore resample repositories with replacement while retaining every aligned instance in each sampled repository. This clustered bootstrap uses the 12 repositories in the SWE-bench Verified-derived arm and the 153 repositories in the random-repository arm. We treat a directional difference as supported by a given analysis only when its interval excludes zero. These intervals quantify uncertainty under instance or repository resampling of the recorded runs; they do not estimate run-to-run variability from repeated stochastic model executions.
	
	\section{Experimental Setup}
	\label{sec:setup}
	
	\subsection{Artifact and Serving Environment}
	All experiments are conducted on a fixed released version of IssueLoc-Bench. For each instance, we use the released task input together with the aligned commit-linked label metadata to reconstruct the corresponding pre-fix repository snapshot. Specifically, evaluation is performed on the repository state immediately preceding the historical human resolving commit. This ensures that explorers localize against the unresolved codebase rather than against post-fix artifacts.
	
	Experiments are orchestrated on a server with four NVIDIA RTX 6000 Ada GPUs. Models are served locally through vLLM OpenAI-compatible endpoints. Every model receives the same localization task, read-only interaction contract, and per-instance time budget.
	
	\subsection{Models and Interactive Runner}
	We evaluate the following five checkpoints:
	\begin{itemize}
		\item \path{zai-org/GLM-4.7-Flash};
		\item \path{google/gemma-4-E4B-it};
		\item \path{Qwen/Qwen3-30B-A3B-Instruct-2507};
		\item \path{Qwen/Qwen3-Coder-30B-A3B-Instruct}; and
		\item \path{Qwen/Qwen3-4B-Instruct-2507}.
	\end{itemize}
	
	These checkpoints cover general-purpose, code-specialized, and compact explorer configurations. The controlled comparison asks how model choice changes explorer quality and observed operational cost.
	
	\label{sec:setup-mini}
	Our setting uses an interactive mini-agent localization runner operating over the materialized pre-fix snapshot. Concretely, this runner follows a constrained mini-SWE-agent-style interaction pattern: for each instance, it prepares a temporary workspace, constructs the issue prompt, and launches an agent that is instructed to perform localization only and terminate with a final structured ranked file prediction.
	
	The agent interacts with the repository exclusively through shell-based read-only inspection commands. In our standard configuration, prompts restrict tool use to commands such as \texttt{ls}, \texttt{find}, \texttt{grep}, \texttt{cat}, \texttt{nl}, and \texttt{sed -n}, preventing file modification and keeping the interaction focused on repository inspection rather than repair. In the standard setup, the per-instance timeout is 600 seconds.
	
	\subsection{Implementation and Evaluation Procedure}
	\label{sec:setup-infra}
	The released artifact contains the components required to reconstruct and evaluate the reported experiments: deterministic snapshot materialization, constrained per-instance execution, the shared evaluator, runnable localization pipelines, and result-generation utilities. Runs use isolated workspaces and support optional Docker sandboxing, parallel and resumable execution, and persistent interaction traces. These traces record elapsed time, interaction counts, token-usage metadata, and execution errors, which are used to analyze localization quality jointly with operational behavior.
	
	\label{sec:setup-eval}
	After prediction generation, explorer outputs are scored against the released label manifests using the shared evaluation pipeline. We evaluate the two arms in Table~\ref{tab:evaluation-arms} for each model under the same interactive runner. Unless otherwise specified, the results use the markdown-ignored evaluation view, in which markdown- and text-like files are excluded during scoring so that the analysis focuses on code-file localization. All runs use the same issue-text mode within each arm. For IssueLoc instances this corresponds to title and body fields; for transformed SWE-bench instances it uses linked issue text when available and the SWE-bench problem statement as fallback.
	
	\section{Results}
	\label{sec:results}
	The results focus on explorer-model allocation under the fixed interactive interface. Tables~\ref{tab:swe-results} and~\ref{tab:random-results} report raw point estimates, and Table~\ref{tab:paired-differences} reports paired uncertainty. We also express quality retention and cost reduction relative to the highest-quality reference explorer.
	
	\subsection{Evaluation-Arm Results}
	\paragraph{SWE-bench Verified-Derived Arm.}
	Table~\ref{tab:swe-results} shows that the reference explorer leads every localization metric on the SWE-bench Verified-derived arm, but also produces the longest and most token-intensive trajectories. Qwen3-Coder-30B-A3B comes closest on top-three discovery. Qwen3-30B-A3B provides the strongest alternative-explorer F1 and the lowest token total, while Gemma-4-E4B and Qwen3-4B occupy shorter-time operating points.
	
	The ranking columns also show that the comparison is effectively concentrated near the head of the list. Almost every prediction contains no more than three files, so Hit@5 usually equals Hit@3. Only Qwen3-Coder gains an additional hit below rank three. This output pattern reinforces Hit@3 and R@3 as the most informative fixed-budget handoff measures for these runs.
	
	\begin{table*}[t]
		\caption{Explorer quality and operational cost on the SWE-bench Verified-derived arm (499 instances).}
		\label{tab:swe-results}
		\centering
		\scriptsize
		\setlength{\tabcolsep}{2pt}
		\begin{tabular}{lrrrrrrr|rrrrrr}
			\toprule
			& \multicolumn{5}{c}{\textbf{Explorer Utility}} &
			\multicolumn{2}{c}{\textbf{Strict Secondary}} &
			\multicolumn{6}{c}{\textbf{Average Ops}} \\
			\cmidrule(lr){2-6}\cmidrule(lr){7-8}\cmidrule(lr){9-14}
			Model & Hit@1 & Hit@3 & Hit@5 & R@3 & MRR & F1 & Exact & Error (\%) & Steps & Time (s) & Input tok. & Output tok. & Total tok. \\
			\midrule
			GLM-4.7-Flash & \textbf{0.7535} & \textbf{0.7756} & \textbf{0.7756} & \textbf{0.7291} & \textbf{0.7642} & \textbf{0.6965} & \textbf{0.5691} & 6.6 & 27.02 & 241.58 & 230373 & 2921 & 233294 \\
			Gemma-4-E4B & 0.6653 & 0.6774 & 0.6774 & 0.6271 & 0.6710 & 0.6281 & 0.5551 & \textbf{0.6} & 5.27 & 43.09 & 25076 & 1563 & 26640 \\
			Qwen3-30B-A3B & 0.6693 & 0.6934 & 0.6934 & 0.6495 & 0.6814 & 0.6400 & 0.5571 & 4.8 & 4.77 & 66.28 & \textbf{10376} & 748 & \textbf{11123} \\
			Qwen3-Coder-30B-A3B & 0.6673 & 0.7315 & 0.7335 & 0.6816 & 0.6986 & 0.6173 & 0.4449 & 2.2 & 8.34 & 81.29 & 25954 & 1162 & 27116 \\
			Qwen3-4B & 0.6052 & 0.6453 & 0.6453 & 0.5956 & 0.6232 & 0.5711 & 0.4629 & 0.8 & 3.49 & \textbf{30.15} & 14412 & \textbf{564} & 14976 \\
			\bottomrule
		\end{tabular}
	\end{table*}
	
	\paragraph{Random-Repository Arm.}
	Table~\ref{tab:random-results} shows the same broad ordering on our random-repository arm: the reference explorer leads localization quality and uses the most time and tokens. Among the alternatives, Qwen3-30B-A3B has the strongest Hit@3, F1, and exact match, whereas Qwen3-Coder-30B-A3B has the strongest R@3. Gemma-4-E4B and Qwen3-4B again trade additional quality for shorter mean completion time.
	
	\begin{table*}[t]
		\caption{Explorer quality and operational cost on the IssueLoc random-repository arm (500 instances).}
		\label{tab:random-results}
		\centering
		\scriptsize
		\setlength{\tabcolsep}{2pt}
		\begin{tabular}{lrrrrrrr|rrrrrr}
			\toprule
			& \multicolumn{5}{c}{\textbf{Explorer Utility}} &
			\multicolumn{2}{c}{\textbf{Strict Secondary}} &
			\multicolumn{6}{c}{\textbf{Average Ops}} \\
			\cmidrule(lr){2-6}\cmidrule(lr){7-8}\cmidrule(lr){9-14}
			Model & Hit@1 & Hit@3 & Hit@5 & R@3 & MRR & F1 & Exact & Error (\%) & Steps & Time (s) & Input tok. & Output tok. & Total tok. \\
			\midrule
			GLM-4.7-Flash & \textbf{0.7720} & \textbf{0.8400} & \textbf{0.8460} & \textbf{0.6671} & \textbf{0.8058} & \textbf{0.6724} & \textbf{0.4300} & 0.8 & 18.30 & 142.95 & 139941 & 1822 & 141762 \\
			Gemma-4-E4B & 0.6820 & 0.7420 & 0.7440 & 0.5664 & 0.7115 & 0.5787 & 0.3880 & 1.0 & 5.17 & 43.80 & 20441 & 1444 & 21885 \\
			Qwen3-30B-A3B & 0.7080 & 0.7560 & 0.7560 & 0.5819 & 0.7307 & 0.5963 & 0.3920 & 2.6 & 4.70 & 51.65 & \textbf{9876} & 752 & \textbf{10629} \\
			Qwen3-Coder-30B-A3B & 0.6960 & 0.7460 & 0.7500 & 0.5858 & 0.7209 & 0.5730 & 0.3480 & 4.0 & 7.30 & 84.14 & 21765 & 1002 & 22767 \\
			Qwen3-4B & 0.5820 & 0.6540 & 0.6580 & 0.4915 & 0.6164 & 0.4880 & 0.3040 & \textbf{0.6} & 3.29 & \textbf{29.50} & 10558 & \textbf{555} & 11113 \\
			\bottomrule
		\end{tabular}
	\end{table*}
	
	\subsection{Paired Model-Comparison Uncertainty}
	\begin{table*}[t]
	\caption{Paired alternative-minus-reference mean differences with 95\% per-instance bootstrap confidence intervals. Negative operational differences indicate lower cost.}
	\label{tab:paired-differences}
	\centering
	\scriptsize
	\setlength{\tabcolsep}{2pt}
	\begin{tabular}{llrrrrr}
		\toprule
		Arm & Explorer & $\Delta$Hit@3 & $\Delta$R@3 & $\Delta$F1 & $\Delta$Time (s) & $\Delta$Tokens (K) \\
		\midrule
		SWE-derived & Gemma-4-E4B & -0.098 [-0.134, -0.062] & -0.102 [-0.137, -0.066] & -0.068 [-0.103, -0.034] & -198.5 [-212.3, -184.8] & -207.4 [-227.9, -187.7] \\
		SWE-derived & Qwen3-30B & -0.082 [-0.118, -0.046] & -0.080 [-0.114, -0.045] & -0.056 [-0.090, -0.023] & -175.3 [-191.7, -158.8] & -218.9 [-239.5, -199.2] \\
		SWE-derived & Qwen3-Coder-30B & -0.044 [-0.080, -0.008] & -0.048 [-0.083, -0.012] & -0.079 [-0.113, -0.046] & -160.3 [-175.8, -145.0] & -208.9 [-229.4, -189.3] \\
		SWE-derived & Qwen3-4B & -0.130 [-0.170, -0.090] & -0.134 [-0.173, -0.094] & -0.125 [-0.164, -0.088] & -211.4 [-225.8, -197.4] & -218.7 [-242.3, -192.7] \\
		\midrule
		Random repository & Gemma-4-E4B & -0.098 [-0.138, -0.058] & -0.101 [-0.134, -0.067] & -0.094 [-0.128, -0.061] & -99.2 [-110.1, -88.8] & -120.5 [-134.3, -107.7] \\
		Random repository & Qwen3-30B & -0.084 [-0.120, -0.046] & -0.085 [-0.117, -0.054] & -0.076 [-0.108, -0.044] & -91.3 [-102.9, -79.9] & -128.3 [-142.1, -115.0] \\
		Random repository & Qwen3-Coder-30B & -0.094 [-0.136, -0.052] & -0.081 [-0.117, -0.047] & -0.099 [-0.132, -0.066] & -58.8 [-70.9, -46.7] & -117.4 [-131.1, -104.0] \\
		Random repository & Qwen3-4B & -0.186 [-0.232, -0.142] & -0.176 [-0.215, -0.137] & -0.184 [-0.223, -0.147] & -113.5 [-123.5, -103.5] & -131.1 [-147.4, -114.0] \\
		\bottomrule
	\end{tabular}
\end{table*}

	At the instance level, every alternative explorer has Hit@3, R@3, and F1 intervals below zero relative to the reference, while every time and token interval favors the alternative. The smallest SWE-derived Hit@3 gap belongs to Qwen3-Coder; Qwen3-30B has the smallest F1 gap and is also closest on both measures in the random-repository arm. Table~\ref{tab:paired-differences} reports the corresponding effect sizes and intervals. Quality and time use the complete arms; token comparisons use instances with usage recorded for both paired runs.
	
	Repository clustering chiefly changes the SWE-derived interpretation. Its Hit@3 and F1 intervals include zero for Gemma and Qwen3-30B, as does Qwen3-Coder's Hit@3 interval; several R@3 and exact-match contrasts are similarly unresolved. By contrast, all random-arm Hit@3, R@3, and F1 gaps remain below zero. The operational reductions remain supported under clustering on both arms. Thus, the cost separation is stable, while the strength of the quality ordering depends on whether variation is measured across tasks or across repositories.
	
	\subsection{Quality Retention and Cost Reduction}
	We normalize quality and cost against the reference explorer, which has the highest point estimates on both arms. For quality $Q$, retention is $100Q_m/Q_{\mathrm{ref}}$; for operational cost $C$, reduction is $100(1-C_m/C_{\mathrm{ref}})$.
	
	The normalized results expose distinct operating points rather than a single uniformly best alternative. Gemma retains at least 86\% of the reference Hit@3 and F1 in both arms while cutting mean agent time by more than two thirds. Qwen3-30B provides the strongest alternative F1 and reduces observed tokens by more than 90\%. Qwen3-Coder sacrifices more strict set recovery but preserves the most SWE-derived top-three discovery. Table~\ref{tab:retention-cost} presents the complete comparison.
	
	\begin{table*}[t]
		\caption{Quality retention and cost reduction relative to the highest-quality reference explorer in each evaluation arm. Positive reduction means lower observed cost.}
		\label{tab:retention-cost}
		\centering
		\scriptsize
		\setlength{\tabcolsep}{4pt}
		\begin{tabular}{llrrrrr}
			\toprule
			Arm & Explorer & Hit@3 retained & F1 retained & Agent-time reduction & Token reduction & API-cost reduction \\
			\midrule
			SWE-bench Verified-derived & GLM-4.7-Flash & 100.0\% & 100.0\% & 0.0\% & 0.0\% & 0.0\% \\
			SWE-bench Verified-derived & Gemma-4-E4B & 87.3\% & 90.2\% & 82.2\% & 88.6\% & 87.9\% \\
			SWE-bench Verified-derived & Qwen3-30B-A3B & 89.4\% & 91.9\% & 72.6\% & 95.2\% & 80.8\% \\
			SWE-bench Verified-derived & Qwen3-Coder-30B-A3B & 94.3\% & 88.6\% & 66.4\% & 88.4\% & 34.1\% \\
			SWE-bench Verified-derived & Qwen3-4B & 83.2\% & 82.0\% & 87.5\% & 93.6\% & 92.2\% \\
			\midrule
			IssueLoc random-repository & GLM-4.7-Flash & 100.0\% & 100.0\% & 0.0\% & 0.0\% & 0.0\% \\
			IssueLoc random-repository & Gemma-4-E4B & 88.3\% & 86.1\% & 69.4\% & 84.6\% & 83.4\% \\
			IssueLoc random-repository & Qwen3-30B-A3B & 90.0\% & 88.7\% & 63.9\% & 92.5\% & 69.5\% \\
			IssueLoc random-repository & Qwen3-Coder-30B-A3B & 88.8\% & 85.2\% & 41.1\% & 83.9\% & 8.8\% \\
			IssueLoc random-repository & Qwen3-4B & 77.9\% & 72.6\% & 79.4\% & 92.2\% & 90.2\% \\
			\bottomrule
		\end{tabular}
	\end{table*}
	
\begin{figure*}[t]
		\centering
		\includegraphics[width=0.9\textwidth]{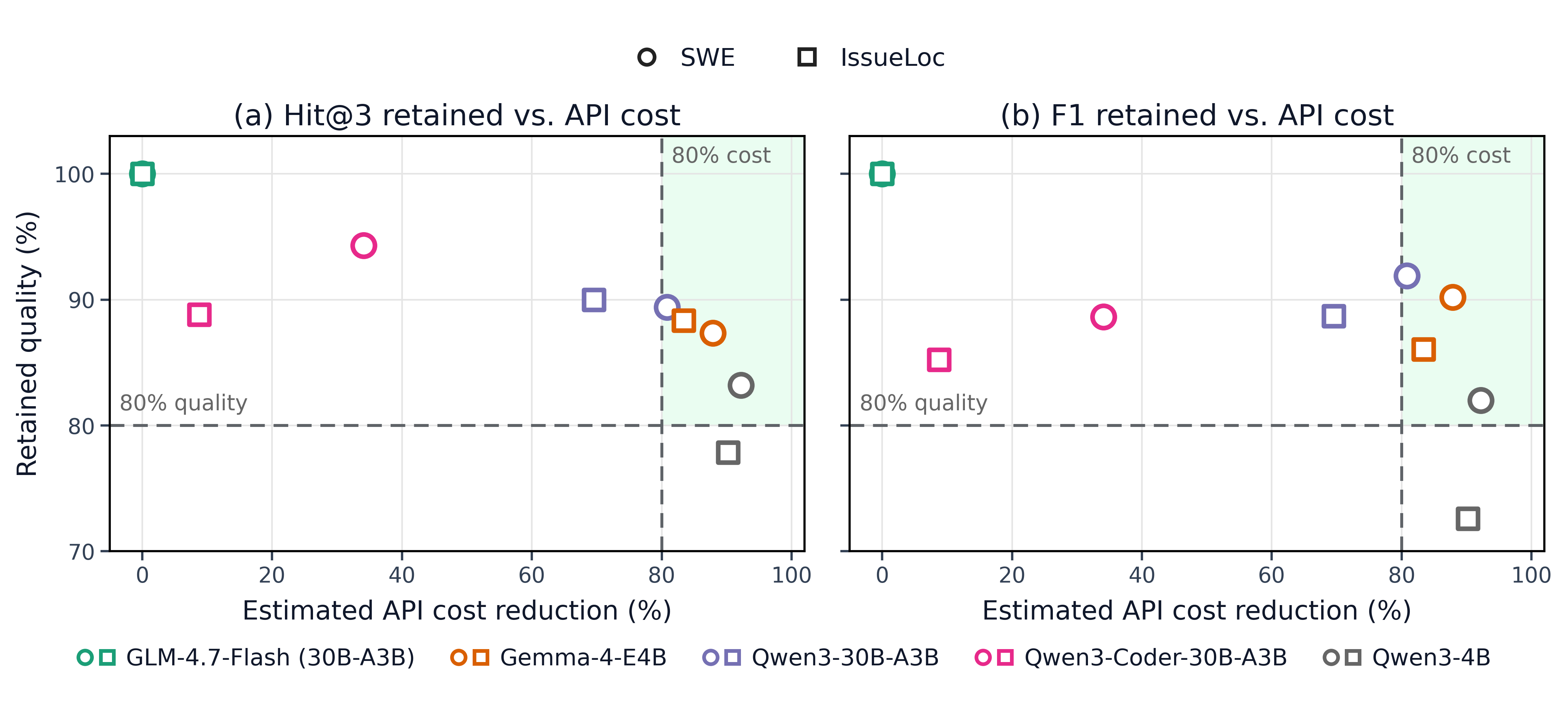}
		\caption{Quality retention versus estimated API-cost reduction. Dashed lines mark the 80/80 operating region.}
		\Description{Two scatter plots comparing retained quality with estimated API-cost reduction for the SWE-bench Verified-derived and IssueLoc random-repository arms. Panel a uses Hit@3 retention and panel b uses F1 retention. Circles mark the SWE-bench Verified-derived arm, squares mark the IssueLoc random-repository arm, colors identify explorer models, dashed lines mark 80 percent estimated cost reduction and 80 percent retained quality, and a shaded upper-right region marks points satisfying both thresholds.}
		\label{fig:cross-arm-api-cost-tradeoff}
	\end{figure*}

	Figure~\ref{fig:cross-arm-api-cost-tradeoff} combines retained quality with API-equivalent cost reduction using the model-specific input and output rates recorded in mid-June 2026. Qwen3-30B remains cost-competitive through shorter, less token-intensive trajectories; Gemma combines low token prices with lower mean agent time, while Qwen3-Coder's output-token price offsets more of its token reduction. The figure complements the hardware-dependent time measurements with a price-based deployment view.
	
	\subsection{Answers to the Research Questions}
	\paragraph{RQ1: Quality Retention.}
	The best-retaining alternative depends on the arm and handoff metric. Qwen3-Coder is strongest for SWE-derived candidate discovery, while Qwen3-30B is strongest for random-arm discovery and strict F1. Thus, retention is not a single property of a checkpoint; it depends on whether the comparison rewards an early entry point, broader top-three coverage, or calibration of the emitted set.

	\paragraph{RQ2: Operational Cost.}
	All alternative explorers reduce time and tokens, with operational intervals remaining below zero under both instance-level and repository-clustered resampling. These reductions describe complete trajectories under the recorded runner and serving configuration, including interaction count, tool use, stopping behavior, and runtime; they are not estimates of intrinsic checkpoint speed.

	\paragraph{RQ3: Handoff Contract.}
	The relative advantage of candidate discovery over strict set recovery is model- and arm-dependent. Qwen3-Coder shows the clearest soft-handoff-oriented trade-off on the SWE-derived arm, retaining the strongest alternative early ranking despite lower F1. Gemma and Qwen3-30B retain strict set performance more strongly, so lower observed cost does not uniformly imply a larger exact-recovery degradation.

	\paragraph{RQ4: Cross-Arm Consistency.}
	The broad quality--cost trade-off recurs across arms, but the ordering among alternatives changes with the metric and arm. Repository clustering also weakens several SWE-derived quality comparisons, so cross-repository conclusions are more qualified than the corresponding instance-level comparisons.
	
	\section{Analysis and Discussion}
	\label{sec:analysis}
	
	\subsection{Explorer Operating Points and Handoff Contracts}
	The results do not identify a universally best lower-cost explorer; they expose different Pareto-like operating points. The reference configuration maximizes observed localization quality. Qwen3-Coder is attractive when early SWE-derived discovery matters; Qwen3-30B is the most balanced alternative for strict recovery and token usage; Gemma favors shorter trajectories; and Qwen3-4B is the aggressive low-time point. Explorer selection therefore depends on which quality dimension the handoff consumes and which operational resource is constrained.

	R@3 makes this distinction visible. Binary Hit@3 asks whether the explorer supplies at least one historically relevant starting point; R@3 asks how much of a multi-file footprint is already present in that same budget. On the random arm, Qwen3-30B has stronger Hit@3 and F1, while Qwen3-Coder has slightly stronger R@3. A downstream stage that can navigate outward from one plausible implementation file may value the former; one receiving a small, fixed context bundle may value broader initial coverage. Neither preference can be inferred from exact match alone.

	The historical footprint explains why early-ranking metrics are an appropriate proxy when downstream access remains recoverable. Human commits frequently modify a mixture of implementation, tests, configuration, or supporting files. Exact recovery treats every retained file symmetrically, whereas a later stage may use one high-ranked entry point to continue exploring repository structure. A ranked list therefore narrows the initial search space and orders the context to inspect first. F1 and exact match remain the relevant measures when the predicted set is a hard boundary.

	The metric panel should consequently be read as a set of contract-specific views, not as interchangeable measures of one latent score. Hit@3 records whether a small candidate budget contains any historical target, R@3 records how much of the target footprint that budget covers, and MRR rewards earlier placement of the first target. These measures characterize the recoverability of the initial search state, but they do not establish the quality of a subsequent patch. F1 and exact match instead compare the emitted and historical sets, making them better aligned with restrictive context selection. Reporting both families prevents an explorer that is useful for one contract from being selected or rejected using a metric intended for the other.

	This interpretation connects the upstream and downstream evidence without treating stage metrics as repair outcomes. Loc2Repair shows that a lightweight localization signal can improve multiple repair backbones \cite{awad2026loc2repairframeworkevaluatingimpact}; the present results establish distinct quality--cost choices for producing such a signal. Together, they motivate treating localization as a service with an explicit output contract, candidate budget, and escalation rule instead of an inseparable prelude to patch generation.
	
	\subsection{Cross-Arm Generality and Repository Dependence}
	The most important cross-arm result is directional stability. In both settings, the reference configuration provides the highest observed localization quality, and every alternative yields lower time and token use. This recurrence matters because the arms stress different aspects of exploration. The SWE-derived tasks require navigation through substantially larger repository trees, while our random-repository tasks more often require recovering several historically changed files. The same broad trade-off appears under both conditions, so it is not tied only to one repository-size or gold-set regime.
	
	The model ordering within that broad pattern is not identical. Qwen3-Coder is especially competitive on candidate coverage in the SWE-derived arm, whereas Qwen3-30B is the more balanced alternative on the random-repository arm. This variation argues against selecting an explorer from a single pooled score. A deployment dominated by a few large framework repositories may value broad search and strong early ranking; a service spanning many unrelated projects may prefer the model whose balance of discovery and set recovery persists across repository types.
	
	The clustered analysis clarifies what can be generalized from the two arms. Instance resampling asks whether the recorded model difference is stable across issues in the benchmark. Repository resampling asks whether it is stable when entire repositories, rather than individual issues, are treated as the sampling unit. Because the SWE-derived arm contains only a small number of repositories, a few projects can influence its aggregate ordering and clustered intervals are necessarily less decisive. The random arm distributes evidence across many more repositories, and its quality ordering remains clearer after clustering. These are complementary views: task-level intervals describe expected performance over a similar issue mix, while clustered intervals better reflect transfer across repositories.
	
	\subsection{Escalation-Oriented Explorer Design}
	These results motivate an escalation-oriented design in which explorer-model allocation becomes a control policy rather than a one-time choice. A natural soft-handoff policy would begin with a lower-cost explorer, pass its ranked files to the repair stage, and retain the ability to broaden repository access. A stronger explorer could then be reserved for cases in which the first trajectory produces no valid candidates, the candidates fail to support a plausible diagnosis, or later inspection uncovers evidence outside the initial neighborhood. Such a policy could allocate exploration budget conditionally rather than selecting one operating point for every issue.

	The escalation signal would need to correspond to an observable failure of the handoff. Runner errors and empty predictions are immediate signals. A downstream repairer could also report that the supplied files do not contain the described behavior, relevant definitions, or reachable dependencies. For a hard gate, escalation would need to occur earlier because ordinary browsing cannot recover an omitted file. In that setting, F1 and exact match become more important selection criteria, and the gate should provide an explicit escape route rather than silently forcing repair within an incomplete set.

	Candidate budgets are another prospective allocation lever. A small budget limits context transfer and encourages precise ranking, but multi-file issues may require broader coverage. The joint use of Hit@3 and R@3 separates these cases: the former measures whether the budget supplies a starting point, and the latter measures how much of the known footprint fits inside it. A deployment could therefore tune candidate count independently of checkpoint choice. For example, a lower-cost explorer could emit a compact initial list, after which the repairer requests expansion only when cross-file evidence is needed.

	A path ranking also separates \emph{selection} from \emph{context transfer}. The explorer need not copy full files into the repair prompt; the repairer can open candidates in order and fetch related context as evidence develops. A low-cost explorer that forwards a large, undifferentiated prompt merely moves the operational burden downstream. Ranked paths instead form a compact control-plane artifact whose quality can be evaluated independently of the repair model and context-packing strategy.

	Evaluating these policies end to end is an important next step. The present metrics identify plausible explorer operating points and observable escalation signals, but a fixed-repairer experiment is needed to determine their effects on resolved rate and total pipeline cost.

	\subsection{Trajectory Cost and Practical Model Selection}
	The operational differences are shaped mainly by how long each explorer continues searching. The reference explorer takes substantially more interaction steps in both arms, while per-step elapsed time is closer among several models. Its total-time gap therefore reflects a longer exploration policy as well as serving/runtime characteristics. Token consumption likewise combines the number of turns with the repository context carried through them. The observed totals are trajectory costs of complete explorer configurations---checkpoint behavior, tool use, stopping policy, and serving stack---rather than intrinsic measures of model speed.

	This decomposition suggests two independent optimization directions. Model selection changes how effectively each turn converts repository evidence into ranked candidates. Runner policy controls when the explorer has enough evidence to stop. A strong model with an unnecessarily long stopping policy can be expensive, while a compact model that stops too early can miss important cross-file relationships. Reporting steps alongside time, tokens, and localization quality prevents these behaviors from being collapsed into a single efficiency label.

	The fixed interface is important to this interpretation. Every checkpoint receives the same task role, read-only repository access, and output schema, so the comparison holds the interaction contract constant. It does not, however, force checkpoints to take the same number of turns, inspect the same files, or stop after the same evidence. Those behaviors are part of the measured explorer operating point. A deployment choosing among the configurations pays for the resulting trajectory as a whole; isolating model throughput or equalizing turn count would answer a different question.

	The resulting selection rule follows the handoff contract and service objective. The reference configuration has the highest observed quality when that criterion dominates. Qwen3-30B offers the most balanced alternative when strict set recovery and token use both matter. Qwen3-Coder favors top-three discovery and coverage, Gemma favors shorter total agent time while retaining substantial quality, and Qwen3-4B represents the most aggressive time-saving point. No alternative dominates across quality, time, and tokens; the appropriate choice depends on which resource is scarce and whether the downstream stage can recover from an incomplete prediction.

	\section{Threats to Validity}
	\label{sec:threats}

	\paragraph{Historical-Footprint Supervision.}
	Gold files are the filtered historical commit footprint, not a proof that every touched file is necessary or that another valid repair would modify the same set. This makes exact match deliberately strict and gives ranking and coverage metrics an important complementary role. The code-focused view also excludes newly added, removed, binary-like, and entirely text-like targets; conclusions therefore concern localization of existing implementation artifacts.

	\paragraph{Repository and Model Sampling.}
	The SWE-derived arm contains many issues from relatively few repositories; the random arm covers more repositories but has its own IssueLoc linkage and filtering criteria. We therefore avoid pooling the arms and report repository-clustered sensitivity alongside instance-level intervals. Both arms use public repositories, so possible exposure during model training cannot be measured from the artifact. Transfer to private codebases, very large monorepos, other language distributions, and other agent interfaces remains a separate empirical question.
	
	\paragraph{Operational Measurements.}
	Time and token use describe the complete recorded operating points: model, serving stack, tool interaction, stopping behavior, and timeout policy. They are deployment-relevant totals, not isolated measures of model throughput. Token comparisons are pairwise-complete when usage metadata is present for both runs, and the API-cost view applies contemporaneous public prices rather than observed invoices. The paired bootstrap quantifies variation across recorded instances or repositories; repeated executions would be needed to quantify stochastic trajectory variation.
	
	\paragraph{Pipeline Scope.}
	The experiment isolates localization so that explorer choice can be compared under a fixed interface. Its rankings therefore guide stage allocation: end-to-end benefit also depends on the repair model, candidate budget, context format, and escalation policy. Evaluating those components in a factorial explorer--repair study would connect the present stage metrics to resolved rate and total pipeline cost.
	
	\section{Conclusion}
	\label{sec:conclusion}
	
	Empirically, under the fixed read-only interface, the highest-quality explorer leads the localization metrics, while the alternatives provide substantially lower-time and lower-token operating points. IssueLoc-Bench makes this upstream allocation visible through early ranking, top-three gold coverage, historical set recovery, trajectory length, agent time, and tokens across two complementary arms. Paired comparisons establish the task-level trade-offs, while repository-clustered sensitivity appropriately tempers several SWE-derived quality contrasts.
	
	Interpretively, no alternative dominates across ranking quality, strict recovery, time, and tokens. The preferred explorer depends on how localization is consumed: Hit@3, R@3, and MRR characterize a recoverable candidate handoff, whereas F1 and exact match characterize a restrictive file gate. Architecturally, these results support treating repository exploration as a separately measurable and budgetable stage in modular coding-agent pipelines, with an explicit output contract and observable trajectory cost.
	
	\paragraph{Future Work.}
	Determining whether these localization trade-offs translate into end-to-end repair trade-offs requires a fixed-repairer evaluation and is left to future work. Such an evaluation can vary the explorer, top-$k$ budget, and escalation rule while measuring resolved rate, recovery beyond the initial handoff, and total pipeline cost. Further directions include adaptive candidate budgets, symbol- and test-level handoffs, temporally filtered issue context, private-repository evaluation, and supervision that distinguishes minimal causal files from the broader historical commit footprint.
		
	\section{Data Availability Statement}
	The artifact at \url{https://github.com/mohammad-nour-alawad/IssueLoc-bench} contains task and label manifests, the parent-commit snapshot builder, shared evaluator, explorer runners, run outputs, and statistical reporting code. Stable identifiers and traces support reconstruction and rescoring.
	
\begin{acks}
		This work is supported by the Ministry of Economic Development of the Russian Federation (IGK 000000C313925P4C0002), agreement No139-15-2025-010.
\end{acks}
	
	\bibliographystyle{ACM-Reference-Format}
	\bibliography{software}
		
\end{document}